\documentclass[twocolumn,showpacs,aps,pra,longbibliography,superscriptaddress]{revtex4-1}
\usepackage[bookmarks=false,linkcolor=blue,urlcolor=blue,colorlinks,citecolor=blue]{hyperref}

\usepackage{ifthen}
\newboolean{captionSection}
\setboolean{captionSection}{false}
\newboolean{preprint}
\setboolean{preprint}{false}

\usepackage{xr}
\pdfoutput=1 

\usepackage{graphicx}
\usepackage{color}
\usepackage[dvipsnames]{xcolor}
\usepackage{hyperref}
\hypersetup{colorlinks=true, linkcolor=Bittersweet,citecolor=Bittersweet, urlcolor=Bittersweet, bookmarksnumbered=true}

\usepackage{amsmath}
\usepackage{amsthm}
\usepackage{amssymb}
\usepackage{wrapfig}
\usepackage[shortlabels]{enumitem}
\usepackage[retainorgcmds]{IEEEtrantools}
\usepackage{tcolorbox}
\usepackage{enumitem}

\usepackage{lineno}

\begin{document}
	
	\author{Doruk Efe G\"okmen}
    \email[]{gokmen@uchicago.edu}
	\affiliation{Institute for Theoretical Physics, ETH Zurich, 8093 Zurich, Switzerland}
    \affiliation{
    James Franck Institute, The University of Chicago, Chicago, IL 60637, USA}
    \affiliation{
    Department of Statistics, The University of Chicago, Chicago, IL 60637, USA}
    \affiliation{
    National Institute for Theory and Mathematics in Biology, Chicago, IL 60611, USA
    }

	\author{Sounak Biswas}
	\affiliation{Institut für Theoretische Physik und Astrophysik, Universit\"at W\"{u}rzburg, 97074 W\"urzburg, Germany}
	
	\author{Sebastian D. Huber}
	\affiliation{Institute for Theoretical Physics, ETH Zurich, 8093 Zurich, Switzerland}
	
	\author{Zohar Ringel}
	\affiliation{Racah Institute of Physics, Hebrew University, Jerusalem, 9190401, Israel}
	
	\author{Felix Flicker}
    \email[]{flicker@physics.org}
	\affiliation{School of Physics, Tyndall Avenue, Bristol, BS8 1TL, United Kingdom}
	
	\author{Maciej Koch-Janusz}
    \email[]{maciej@haiqu.ai}
	\affiliation{Department of Physics, University of Z\"urich, 8057 Z\"urich, Switzerland}
	\affiliation{
    James Franck Institute, The University of Chicago, Chicago, IL 60637, USA}
    \affiliation{Haiqu, Inc., 95 Third Street, San Francisco, California 94103, USA}

    \title{Compression theory for inhomogeneous systems}
	
	\begin{abstract}
        The physics of complex systems stands to greatly benefit from the qualitative changes in data availability and advances in data-driven computational methods. Many of these systems can be represented by interacting degrees of freedom on inhomogeneous graphs. However, the lack of translational invariance presents a fundamental challenge to theoretical tools, such as the renormalization group, which were so successful in characterizing the universal physical behaviour in critical phenomena. Here we show that compression theory allows the extraction of relevant degrees of freedom in arbitrary geometries, and the development of efficient numerical tools to build an effective theory from data. We demonstrate our method by applying it to a strongly correlated system on an Ammann-Beenker quasicrystal, where it discovers an exotic critical point with broken conformal symmetry. We also apply it to an antiferromagnetic system on non-bipartite random graphs, where any periodicity is absent.
	\end{abstract}
	
	\maketitle
	
\section*{Introduction}	
Dramatic improvements in data availability, stemming from both experiments and simulation, have enabled the exploration of increasingly complex physical systems. 
A glut of raw data does not, however, equate understanding, particularly when its processing easily exceeds our computational resources.
A key objective is to distil data into a succinct theory in terms of appropriate collective variables that uncover and summarize the essence of the system. 
Renormalization Group (RG) approaches in statistical physics provide a systematic path towards that goal~\cite{PhysRevB.4.3174, PhysicsPhysiqueFizika.2.263}.
However, both identifying the relevant degrees of freedom (DOFs), as well as executing the mathematical procedure deriving the effective theory~\cite{PhysRevLett.43.1434, IGLOI2005277} are often very challenging in inhomogeneous systems when prior intuition is scarce.

Many complex systems are inhomogeneous, owing their properties precisely to the lack of translational invariance. Problems as disparate as biological tissue mechanics~\cite{FLETCHER20142291, doi:10.1098/rstb.2015.0520, FARHADIFAR20072095} and properties of metallic glasses~\cite{BERNAL_1960, osti_914160}
can be cast as statistical mechanical problems on irregular graphs. 
The ultimate goal would be a generic understanding of the emergent properties of such systems, much as RG provided an understanding of critical phenomena in translation-invariant systems, 
where the proxy of wavelength can be used to organise the modes to target the relevant low-energy operators~\cite{PhysRevB.4.3184}. While one can still implement scale transformations to perform RG for inhomogeneous systems, the lack of a clear proxy like wavelength makes it necessary to carefully craft the coarse graining locally, particularly since real-space RG can be ill-defined under certain poor choices~\cite{RN274}.

Here we tackle the challenge of inhomogeneity in complex systems with vast configuration spaces.
Formulating the RG of an inhomogeneous system as a lossy compression of information~\cite{infbottle1, gordon2020relevance} on a graph allows us to define the procedure in a geometry-independent manner. This key step is based on the observation that the compression theoretic RG some of us introduced for lattice systems~\cite{gokmen2021statistical, PhysRevX.10.011037, Koch-Janusz2018} can be mathematically generalized to arbitrary graphs, yielding RG informed both of the interactions and of the spatial relations, thus overcoming a major conceptual challenge. The numerical execution of this data-driven procedure entails the difficult task of estimating mutual information for large-dimensional random variables on graphs. This is achieved by using recent advances in machine learning, formulating it as a classification task of distinguishing jointly sampled pairs of variables from those sampled independently (see Methods), \emph{i.e.}~using contrastive learning~\cite{belghazi18a, Poole2019, oord2019}, and extending the computational tools of Ref.~\cite{gokmen2021phase, rsmine_code}.

    \begin{figure*}
        \includegraphics{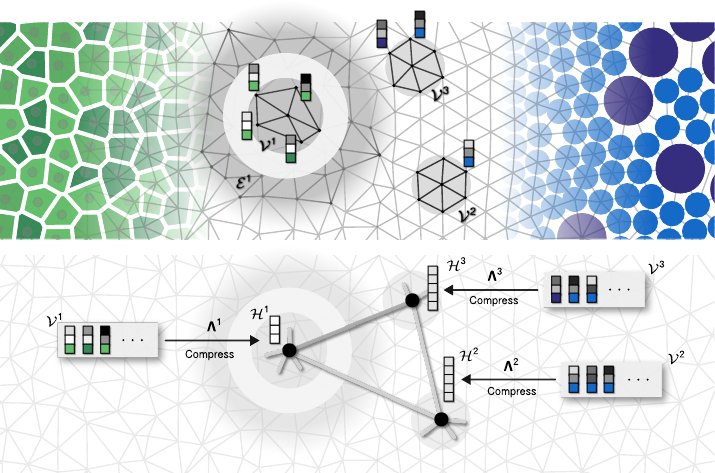}
        \caption{{\bf Schematic for constructing collective degrees of freedom in inhomogeneous systems.} 
            Distinct systems like tissues (left, in green) and colloidal suspensions (right, in blue) can be abstracted into a set of vector degrees of freedom $\mathcal{V}^i$ (indicated by stacks of squares, $i=1,2,3$) living on an irregular graph with local structure. The final component of each vector is shown by a coloured box to indicate potentially different types of internal degree of freedom, unique to each sub-system. To derive a compressed representation of such systems, it is essential to tailor the coarse graining transformation $\boldsymbol{\Lambda}^i$ for each local neighbourhood $i$. This is achieved by an information theoretic variational principle, where $\boldsymbol{\Lambda}^i: \mathcal{V}^i \mapsto \mathcal{H}^i$ maximises the mutual information $I\left(\mathcal{H}^i:\mathcal{E}^i\right)$. This allows the compressed variables $\mathcal{H}^i$ to capture the emergent long-range physics according to the statistics of the surrounding distant environment $\mathcal{E}^i$. Local optimisation can produce compressed variables with varying cardinality across the system, here illustrated by vectors $\mathcal{H}^i$ with varying numbers of components. The connectivity of the emergent \emph{supergraph} is determined through the correlations of the new variables.
        }
        \label{fig:fig1}
    \end{figure*}

Our algorithm assumes the system is defined on a graph $G$ and explicitly constructs new effective DOFs $\mathcal{H}^i$ from local configurations of DOFs $\mathcal{V}^i$ supported on vertices or edges of local subgraphs $V^i$ of $G$.
This is achieved by a coarse graining transformation, which we parametrise as a linear neural network with a set of parameters $\boldsymbol{\Lambda}^i$, and a non-linear discretisation map $\tau$:
\begin{equation}
    \mathcal{H}^i = \tau \left(\boldsymbol{\Lambda}^i \cdot \mathcal{V}^i \right).
\end{equation}
This coarse-graining is optimised locally in region $i$ to maximise the mutual information $I(\mathcal{H}^i:\mathcal{E}^i)$ with the configurations supported on the spatial environment $E^i$ of $V^i$ (see Fig.~\ref{fig:fig1} and Methods)~\cite{gokmen2021statistical, PhysRevX.10.011037, Koch-Janusz2018}:
\begin{equation}\label{eq:rsmi}
    \boldsymbol{\Lambda}^i := \arg\max I(\mathcal{H}^i:\mathcal{E}^i). \\ 
\end{equation}

The disjoint spatial environment subgraph $E^i$ is defined using the graph distance (see Fig.~\ref{fig:fig1}, and Fig.~S1), and its separation from the block $V^i$ has a crucial role of screening out the irrelevant short-range correlations.
The mutual information
\begin{equation}
    I(\mathcal{X}:\mathcal{Y})=H(\mathcal{X}) + H(\mathcal{Y}) - H(\mathcal{X}, \mathcal{Y}),
\end{equation}
quantifies the amount of information the random variable $\mathcal{X}$ reveals about the other $\mathcal{Y}$, by measuring their overlapping contributions to the total entropy $H$.

The maximisation is performed under the constraint that the coarse-grained variable $\mathcal{H}^i$ is subject to a discretisation $\tau$ (see Methods).
This crucial aspect of the variational principle renders the coarse-graining a \emph{lossy} compression map, \emph{i.e.} $I(\mathcal{H}^i:\mathcal{E}^i) < H(\mathcal{V}^i)$, where $H$ is the entropy. Thus, only the most relevant collective features of $\mathcal{V}^i$ that survive this information bottleneck will be stored in $\mathcal{H}^i$. Note that they depend on both the topology of the graph $V^i$ and the interactions of the DOFs $\mathcal{V}^i$ on it, and  information about both is contained in the statistics of the samples $(\mathcal{V}^i,\mathcal{E}^i)$.

The variational principle in Eq.~\eqref{eq:rsmi} provides a powerful substitute for heuristic approaches. Specifically, rather than guessing an important local collective property for coarse graining, the collective DOFs are instead designed by the statistics of their environments.
This is essential for moving beyond translation invariance.
While the coarse graining $\boldsymbol{\Lambda}^i$ erases microscopic fluctuations, its local optimisation allows it to retain the distinct qualitative characteristics that emerge in different spatial regions across an inhomogeneous system. 
This can even be reflected in a non-uniform cardinality of compressed variables $\mathcal{H}^i$, as illustrated in Fig.~\ref{fig:fig1}.
Thanks to the recently discovered links between RG relevance and compression theory~\cite{gordon2020relevance}, this procedure eliminates proxies like wavelength (or even energy, in purely entropic systems) and grants direct access to the operators with low-scaling dimension~\cite{gokmen2021statistical}.
    
We illustrate the power of this approach by explicitly constructing emergent DOFs to confirm an open conjecture regarding the presence of discrete scale invariance (DSI)~\cite{Sornette1998discrete,Mohammad2020,PhysRevE.107.024137} of a strongly correlated statistical model on quasicrystals~\cite{Lloyd_2022}.
This problem is perfectly suited to our method: the combination of constraints and aperiodicity provides a serious challenge for human intuition~\cite{Flicker2020,Lloyd_2022,Singh23}; yet efficient algorithms exist that can generate huge data sets for training machine intuition~\cite{biswas2023discrete}.

Though  quasicrystals lack translational invariance, they possess long-range order~\cite{QC_Senechal,tilingsandpatterns}. 
To show the generality of our method, and its independence of the existence of any special tilings, we also applied it to frustrated antiferromagnets on non-bipartite random graphs lacking any (quasi)periodicity, where it finds the optimal bipartitioning (Section IV of Supplementary Information).
Our graph-based tool leverages the computational back-end of the real-space mutual information neural estimation (RSMI-NE) package~\cite{rsmine_code, gokmen2021phase}, allowing to efficiently explore large-scale phenomena.

	\begin{figure}
		\includegraphics{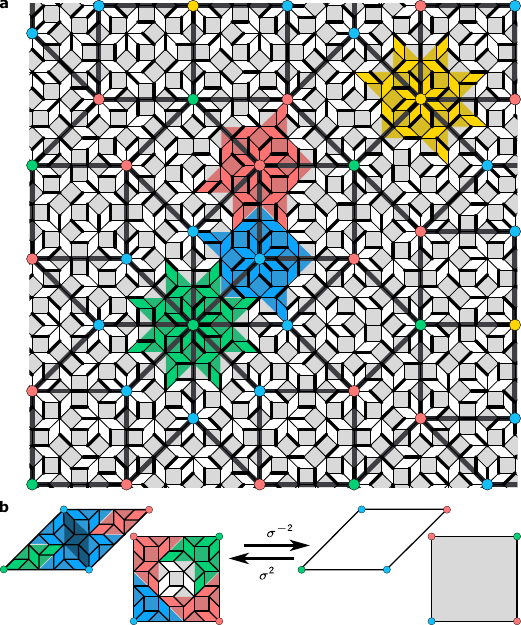}
		\caption{{\bf Self-similarity of the Ammann-Beenker tiling, and the coarse graining blocks.}
			{\bf a} A microscopic dimer configuration on the AB tiling's edges, with an overlaid AB \emph{super-quasilattice}, self-similar to the microscopic one. 
            The effective degree of freedom at a supervertex with valence $n$ will be obtained by coarse graining the dimer configuration in the surrounding polygon tile $V^n$. In total there are 4 classes of such polygons, here shown in green, blue, red and yellow for $n=8,3,4,5$, respectively.
            The shape of the block tile is dictated by the valence $n$ of the central supervertex in matching colour.
			{\bf b} The inflation (deflation) $\sigma^{2(-2)}$ of the elementary rhombi and squares generating the tiling, with parts of the polygonal domains indicated in colour. Coarse graining all such polygonal patches executes a deflation $\sigma^{-2}$ of the original AB quasilattice, yielding the super-quasilattice shown.
        }
        \label{fig:fig2}
	\end{figure}

\section*{Results}
	
\subsection*{Dimers on the Ammann-Beenker tiling} 
The Ammann-Beenker (AB) construction gives quasiperiodic tilings of the plane using two distinct plaquettes: a rhombus and a square~\cite{beenker,QC_Senechal,tilingsandpatterns}. 
Like their famous cousins, the Penrose tilings~\cite{PenroseOriginal}, AB tilings feature diffraction patterns exhibiting crystallographically `forbidden' symmetries, here 8-fold~\cite{QC_Senechal}. Likewise, they can also be generated by a recursive procedure in which an \emph{inflation} map $\sigma$ acts on a small seed patch by composing the constituent plaquettes as shown in Fig.~\ref{fig:fig2}b, and subsequently rescaling edge lengths by the silver ratio $\delta=1+\sqrt{2}$. A special role is played by 8-fold coordinated vertices: under inflations all lower coordinated vertices ultimately become (and stay) 8-vertices. Each 8-vertex is characterised by an \emph{order}, \emph{i.e.}~the maximal number of inverse \emph{deflations} $\sigma^{-1}$ after which it still remains 8-fold coordinated. The order of an 8-vertex intuitively specifies the maximal size of the local 8-fold symmetric patch centred on it.
The quasiperiodic AB lattice (`quasilattice') is thus invariant under discrete rescalings. Such discrete rescalings are easily visualized for even order deflations $\sigma^{-2n}$ by drawing a super-quasilattice connecting 8-fold vertices (Fig.~\ref{fig:fig2}a).
	
In this setting, we consider a dimer model. Dimer models enjoy a deceptively simple definition: microscopic degrees of freedom live on the links of a graph (here, the edges of the quasilattice), which can be either occupied or empty. The key element is a hard local constraint: at every vertex where the links meet, one and only one of the links is occupied by a dimer. This gives rise to a surprisingly rich phenomenology.
Dimer models on regular lattices have been studied extensively, originally due to their purported relevance to high-$T_c$ superconductivity~\cite{PhysRevLett.61.2376}. They have since been shown to support topological order and fractionalisation~\cite{PhysRevLett.86.1881,PhysRevB.65.024504} and exotic critical points~\cite{doi:10.1126/science.1091806}. The classical problem is closely related to the quantum one~\cite{PhysRevLett.94.235702,PhysRevE.74.041124} and has deep connections to combinatorics~\cite{whatisadimer,KASTELEYN19611209,10.2307/2646174} and the study of random surfaces~\cite{dimersamoebae,limitshapes}.
	
Recent work has begun to explore the interplay of (strongly-correlated) dimer physics and quasiperiodicity. Particularly, AB tilings, in contrast to Penrose tilings~\cite{Flicker2020}, host perfectly matched dimer configurations (\emph{i.e.}~with a vanishing density of uncovered vertices) in the thermodynamic limit. Numerically computed dimer correlations exhibit a quasi power-law decay with a complex spatial structure~\cite{Lloyd_2022}. Moreover, the combinatorial proof of perfect matching pointed to a hierarchy of self-similar effective matching problems at different scales.

\begin{figure*}[hbt!]
    \includegraphics{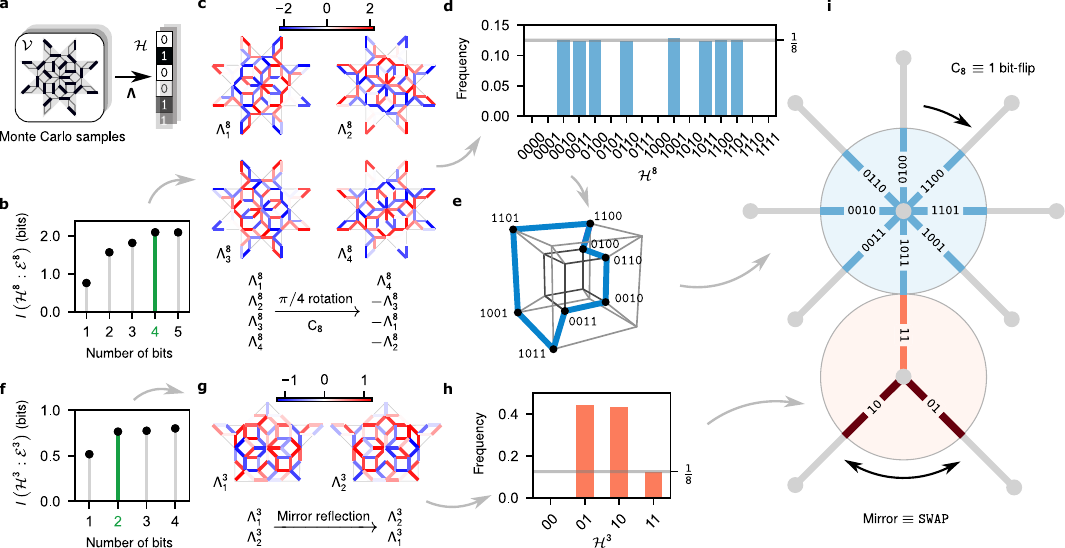}
    \caption{{\bf Finding collective clock variables.}
        {\bf a} Coarse graining transformation $\boldsymbol{\Lambda}$ compressing Monte Carlo configurations $\mathcal{V}$ into bitstrings $\mathcal{H}$ on supervertices of the $\sigma^{-2}$ deflated AB tiling. Each bit $\mathcal{H}_k$ is decided by the sign of the linear transformation $\Lambda_k \cdot \mathcal{V}$. 
        {\bf b}({\bf f}) The length of the bitstring $\mathcal{H}^{8(3)}$ is determined by the saturation point (shown in green) of mutual information at 4 (2) bits at 8- (3-)supervertices. 
        {\bf c}, {\bf g} The respective optimal filters $\boldsymbol{\Lambda}^8$ and $\boldsymbol{\Lambda}^3$ carry a representation of the local spatial symmetries of corresponding supervertices, namely $\mathsf{C}_8$ and mirror. 
        {\bf d}({\bf h}) The probability distributions $P(\mathcal{H}^{8(3)})$ occupy the space of codes sparsely, and form abstract $\mathbb{Z}_{8(3)}$ clock variables. 
        {\bf e} In particular, $\mathcal{H}^8$ forms a closed 8-loop, where each state has exactly two neighbours with Hamming-distance 1. 
        {\bf i} The representations of the local symmetries on filters induce transitions between adjacent clock-states, enabling the identification of abstract clock-states with spatial directions along the links of the quasiperiodic lattice.
    }
    \label{fig:fig3}
\end{figure*} 
	
Taken together, these facts suggest a conjecture that not only the AB tilings themselves, but crucially also the physics of the dimers on the AB tilings, exhibit DSI~\cite{Lloyd_2022} -- a potentially striking and unusual example of the relevance of quasiperiodicity for critical behaviour. A proof and a microscopic physical mechanism at the level of the dimer ensemble has, however, proven elusive.
The putative criticality naturally calls for an RG analysis, but general RG approaches for quasiperiodic systems in $D \geq 2$ dimensions are in their infancy.

In the following, we will first identify the natural block structures to coarse grain the dimer DOFs with a certain scale transformation of the quasilattice.
We will then use our compression approach based on Eq.~\ref{eq:rsmi} to address two key questions regarding the dimer model on AB tilings: What are the collective coarse-grained DOFs, and what is the structure of their correlations?
Finally, by analysing the compressed data provided by our algorithm, we demonstrate the presence of DSI in the dimer model on the AB tiling.

\subsection*{Collective $\mathbb{Z}_n$ clock degrees of freedom}
To construct the collective DOFs, we first need to specify the block regions $\mathcal{V}$ to be coarse grained. In the AB tiling there are natural choices, set by the recursive structure of the AB quasilattice itself~\cite{PhysRevLett.92.047202}. At each scale, the AB tiling can be covered by four `classes' of blocks~\cite{Lloyd_2022} $V^n$, shown in different colours in Fig.~\ref{fig:fig2}, each deflating to vertices of differing connectivity $n$ in the super-quasilattice. In the following, we label each class by the corresponding connectivity $n$ of the super-quasilattice. Our method does not rely on any fine-tuned choice of the block shape (see Section III in Supplementary Information).

In each different class, the algorithm identifies the collective DOF as a $\mathbb{Z}_n$ \emph{clock variable}~\cite{clock_model}, with $n$ the connectivity, or class, of the central supervertex of $V^n$. This is revealed by a variational compression map $\boldsymbol{\Lambda}^n$, which assigns to a Monte Carlo dimer configuration $\mathcal{V}^n$ a short binary code $\mathcal{H}^n$ (Fig.~\ref{fig:fig3}a). 
The binary digits are set by applying individual components $\Lambda^n_k$ to $\mathcal{V}^n$ (itself a long bitstring of dimer occupations in the block). Each component of the vector $\boldsymbol{\Lambda}^n$ is \emph{a priori} a general nonlinear map, parametrized by a NN, whose output is finally binarized. 
	
The length of the code is not supplied, but it is inferred by sequentially increasing the number of components in $\boldsymbol{\Lambda}^n$, and training the compression of $\mathcal{V}^n$ to optimally preserve the mutual information with its environment $\mathcal{E}^n$. Crucially, the maximal retained information about $\mathcal{E}^n$ plateaus with a different optimal code-length depending on the class $\mathcal{V}^n$. 
Particularly, for $V^8$ (green 8-star patch in Fig.~\ref{fig:fig2}) the optimal number of components is four, while for $V^3$ (blue patch in Fig.~\ref{fig:fig2}) it is only two (Fig.~\ref{fig:fig3}b,f). Further, nonlinearity of $\boldsymbol{\Lambda}$ networks does not improve compression: the same amount of information is preserved with only linear components. Optimal linear maps on the space of dimer configurations $\mathcal{V}^n$ on subgraph $V^n$ are shown for $n=8, 3$ in Figs.\ref{fig:fig3}c and g, respectively. 

To unravel the physical content of these encodings, we query the outputs of our algorithm. The code statistics in Fig.~\ref{fig:fig3}d reveal striking features: In class-8, of the sixteen 4-digit binary codes, only eight are ever assigned to $\mathcal{H}^8$, with half of the codes unused. Yet a 3-digit binary encoding, which has exactly eight available codes, is sub-optimal (Fig.~\ref{fig:fig3}b). Moreover, all eight codes have identical probability $1/8$ (Fig.~\ref{fig:fig3}d). This is in contrast to what happens in class-3, where the optimal compression uses three 2-digit binary codes, where only two codes are equiprobable, and the remaining one has probability $1/8$ (Fig.~\ref{fig:fig3}h). These puzzling results indicate that the optimal compression finds structure beyond merely the number of states of the DOF, which is essential to correlations with $\mathcal{E}$, and which is impossible to encode using fewer bits.

The used codes (that is, the $n$ states of effective DOF $\mathcal{H}^n$) are not arbitrary, but are related to the local symmetries of the super-quasilattice.
This can be seen by investigating the structure of codes and the $\boldsymbol{\Lambda}^n$ maps. Notice, for example, that the eight 4-bit states of $\mathcal{H}^8$ can be arranged on a closed 8-cycle, such as (Fig.~\ref{fig:fig3}e) 
\begin{equation}\label{eq:8loop}
    \text{\small ${\tt 1101} \to {\tt 1100} \to \cdots \to {\tt 1001} \to {\tt 1101}$},
\end{equation}
where each code has exactly two 1-bit-distant neighbours.
Interestingly, this solves the four dimensional `coil in the box' problem familiar from coding theory.
Together with the equiprobability of these codes in Fig.~\ref{fig:fig3}d, this cyclic structure hints at a symmetry.
	
Indeed, a class-8 patch $V^8$ of the AB quasilattice is locally symmetric under $\pi/4$ rotations. 
The mutual information should be invariant under the action of this symmetry on the compression map $\boldsymbol{\Lambda}^8$.
Such rotations transform the linear filters $\Lambda^8_k$ via a permutation and inversion of the components, as can be verified visually in Fig.~\ref{fig:fig3}c:
    \begin{equation}\label{eq:c8_rot}
        \mathsf{C}_8:\,\,\,(\Lambda^8_1,\Lambda^8_2,\Lambda^8_3,\Lambda^8_4) \rightarrow (\Lambda^8_4,-\Lambda^8_3,-\Lambda^8_1,-\Lambda^8_2),
    \end{equation}
which is a representation of a generator of the cyclic group $\mathsf{C}_8$. 
We emphasize that it is now the compression map, and consequently the collective DOF now carrying a representation of what is \emph{a priori} a (local) symmetry only of the AB tiling. 

Strikingly, when we apply the transformation Eq.~\eqref{eq:c8_rot} to the states of the coarse grained DOF $\mathcal{H}^8$ (where it now amounts to the permutation of the binary digits and bit-flips), we find that it generates exactly the 8-cycle of Eq.~\eqref{eq:8loop}. 
Since the 1-bit-flip transitions on this cycle are directly induced by $\pi/4$-rotations, the eight states of $\mathcal{H}^8$ can be aligned with the spatial orientations along the eight links of the 8-supervertex, as shown in Fig.~\ref{fig:fig3}i.
This establishes $\mathcal{H}^8$ as a $\mathbb{Z}_8$ \emph{clock variable}.

A similar analysis can be performed for other classes of $V^n$, which have a mirror symmetry. In particular, under the mirror reflection of the class-3 patch $V^3$, the two digits of $\mathcal{H}^3$ are swapped as (see Fig.~\ref{fig:fig3}g)
    \begin{equation}
        {\rm Mirror}:\,\,\, (\Lambda^3_1,\Lambda^3_2) \rightarrow (\Lambda^3_2,\Lambda^3_1).
    \end{equation}
Since the mirror axis is along the edge connecting the 8- and 3-vertices (see Fig.~\ref{fig:fig3}i), it associates the swap-invariant state {\tt 11} with the edge pointing towards the 8-vertex, and the remaining equiprobable states {\tt 01} and {\tt 10} with the other two edges.
Like in class-8, its transformation under symmetries establishes $\mathcal{H}^3$ as a 3-state \emph{clock variable}, whose states can be identified with the super-quasilattice edges. 

Hence, we see that the DOFs of the dimer system remain discrete under coarse graining. 
In particular, we compressed the dimer microstates on the microscopic links into $\mathbb{Z}_n$ clock variables that live on the \emph{vertices} of the underlying super-quasilattice, where they mimic the local symmetries. 
We found that this result holds equally at both $\sigma^{-2}$ and $\sigma^{-4}$ scale transformations, providing the first indication of a DSI.
The persistent discreteness of the collective variables is to be contrasted with the situation on periodic lattices, such as the dimer coverings of the square lattice, which has an emergent continuous U(1) symmetry~\cite{PhysRevB.65.024504}.

\begin{figure*}
    \includegraphics{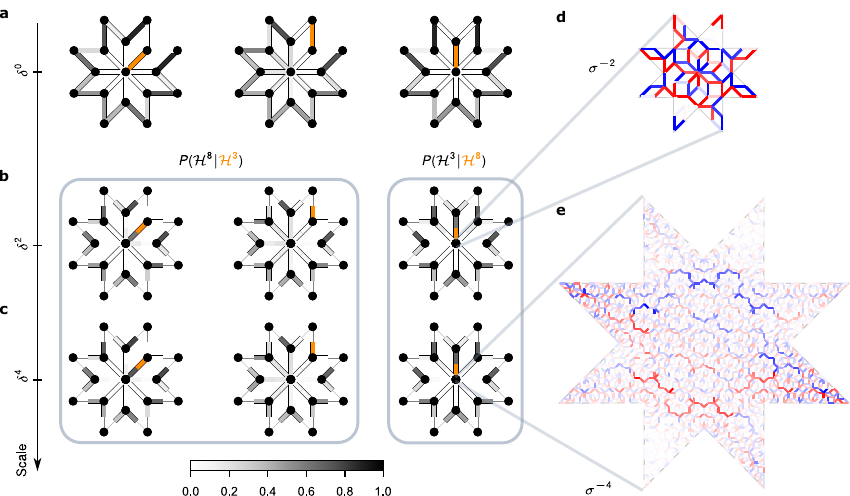}
    \caption{{\bf Emergent dimer exclusion rule and self-similar dimer-dimer correlations across scales.}
        {\bf a} The probability distribution of microscopic (\emph{i.e.~}$\delta^0$) dimers (in greyscale) on an AB patch, conditioned on one of the links (in orange) hosting a dimer.
        {\bf b}, {\bf c} First two columns: the probabilities $P(\mathcal{H}^8|\mathcal{H}^3)$ of the emergent clock variables on the $\delta^{2}$ and $\delta^{4}$ super-quasilattice (in greyscale), conditioned on two distinct states of one of the 3-clocks (in orange). The third column shows distributions $P(\mathcal{H}^3|\mathcal{H}^8)$ conditioned on a state of the central 8-clock. 
        Binding of adjacent clock variables into super-dimers obeying dimer exclusion constraints is revealed by sharply peaked conditional distributions. The effective super-dimers reproduce also longer-range dimer-dimer correlations at both $\delta^2$ and $\delta^4$ scales.
        {\bf d}, {\bf e} Examples of optimal coarse-graining filters producing the central 8-state clock variable at scales $\delta^{2}$ and $\delta^{4}$. The latter comprises $2760$ microscopic links.
    }
    \label{fig:fig4}
\end{figure*}

\subsection*{Binding of clock variables into emergent super-dimers}
Having identified the collective clock DOFs in different classes of blocks individually, we now turn to their correlations, where DSI is manifested fully.
To this end, we simultaneously coarse grain dimer configurations in multiple blocks. 
Deflating the canonical blocks $V^n$ using trained compression maps (Fig.~\ref{fig:fig3}c,g), the correlations of the collective variables $\mathcal{H}^n$ indicate that the effective renormalised model of the clocks has hard-core attractive and repulsive interactions along the links of the AB super-quasilattice (see the bold edges in Fig.~\ref{fig:fig2}a).

We probe the correlations by conditioning on the state of one of the vertices. 
In Fig.~\ref{fig:fig4}b,c, fragments of $\sigma^{-2}$ and $\sigma^{-4}$ super-quasilattices are shown, with the state of the conditioning variable, identified with a direction, in orange, and the conditional distribution of DOFs at the other vertices in grey. Remarkably, this distribution is very strongly correlated, effectively forcing occupation of some states, and excluding others. To wit, when the 3-vertex DOF points towards the 8-vertex, the distribution $P(\mathcal{H}^8|\mathcal{H}^3)$ of the latter is sharply peaked in the matching direction, while no other neighbour of the 3-vertex points towards it (allowing, for example, the identification of the 8-vertex code {\tt 1011} with a specific spatial orientation in Fig.~\ref{fig:fig3}i). Conversely, when the 3-vertex DOF points towards one of its other neighbours, it is `matched' by the latter, while the 8-vertex DOF distribution has zero weight precisely and only in the direction towards that 3-vertex.
	
Examining all such correlations we arrive at a striking conclusion: the effective DOFs in $\mathcal{V}$s throughout the quasilattice are paired with one and only one of their neighbours into emergent \emph{super-dimers} on the edges of the super-quasilattice. The exclusion of certain clock variable orientations in Figs.~\ref{fig:fig3}a-e is a precise reflection of the hard dimer-constraints, which these super-dimers obey. 
Moreover, comparison of further correlations to those of the microscopic dimers in Fig.~\ref{fig:fig4}a reveals that not just the local-dimer constraints, but also longer-range correlations are reproduced correctly.
The physics of the microscopic dimer model on the AB quasilattice is thus replicated, to a high degree of accuracy, at the $\delta^{-2}$ scale and, again, at the $\delta^{-4}$ scale (where `locking' is even sharper, see Fig.~\ref{fig:fig4}c), thereby demonstrating DSI across three scales.

\section*{Discussion}
Guided by the outputs of the RSMI-NE algorithm, we have seen how the quasiperiodicity of the AB quasilattice and the hard-core interactions of the dimer model conspire to \emph{re}create self-similar DOFs at a higher scale, giving rise to DSI.
A parallel work~\cite{biswas2023discrete} gives a microscopic interpretation of the super-dimers as alternating dimer paths with respect to a certain reference configuration, and studies the criticality numerically.

Emergent continuous scale invariance is a standard signature of critical phenomena, being one aspect of conformal symmetry~\cite{cardy2008conformal}. Here we instead encounter an exotic kind of critical phenomenon where this continuous conformal symmetry is broken to a discrete subgroup, thereby complicating the usual effective continuum theory description at large scales.
This is particularly interesting, as it appears to challenge the received wisdom that quasicrystallinity should always be RG irrelevant~\cite{Luck1993}.

We would like to also emphasize the dual computational and conceptual aspect of this result:
In particular for the $\sigma^{-4}$ scale transformation, the RSMI-NE algorithm successfully encodes the symmetries and large-scale correlations in approximately $2^{10^3}$ dimer microstates into a highly structured linear coarse graining map $\boldsymbol{\Lambda}^n$, which is effectively impossible to guess or analyse by human intuition only.
Our demonstration of this approach on an open problem shows how machine learning tools, when paired with a physically motivated objective function, can bridge gaps between complex data and formal physical understanding.

In the Supplementary Information we have also applied the graph RSMI-NE algorithm on a class of non-bipartite random graphs, where we show that it can be used to construct a global order parameter for the frustrated Ising antiferromagnet.
From the dual perspective of combinatorial optimisation, this amounts to solving the well-known optimal graph-bipartitioning problem in combinatorial optimisation and graph theory~\cite{garey1979computers, GAREY1976237, Karp1972, random_bipartitioning}.
These results illustrate the practical applicability of our method  -- formally apparent from its mathematical definition -- to more generic graph topologies, and its independence of the knowledge, or even existence, of preferred spatial blocks.

We therefore conclude that lossy compression allows effective DOFs to be extracted from the structure of information inherent in raw high-dimensional data, and that this approach excels in systems with non-regular geometry.
Given that such cases are the norm rather than the exception in real-world applications, we expect compression theory to become an essential tool in the physical sciences.
%

\section*{Methods}
	\subsection*{Real-space mutual information based coarse graining}
	The method used to construct the effective degrees of freedom (DOFs) is a generalisation of the compression theoretic approach first introduced by some of the authors in Ref.~\cite{Koch-Janusz2018} for translation invariant systems. 
    The key insight is that the random variables, such as the local variable being coarse-grained $\mathcal{V}$ and its environment $\mathcal{E}$, in variational principle in Eq.~\ref{eq:rsmi} can be defined in any metric space, and need not be restricted to regular lattices as in Ref.~\cite{Koch-Janusz2018}. Here we use the graph distance which provides a meaningful measure for spatial length scale on graphs with a local structure. This allows to define the optimal coarse-graining in any geometry, addressing one of the main challenges to applying RG in inhomogeneous systems.

    The above construction is also useful in tackling a second key issue in RG approaches on disordered systems: One has to deal with the changing coarse-grained graph across different scales ~\cite{IGLOI2005277}.
    Here, the effective coarse-grained graph structure is defined by the correlations of the collective DOFs themselves. Thereby the procedure takes account of both the topology of the space and the interactions.
    Endowed with these two properties, the compression principle in Eq.~\ref{eq:rsmi} yields a mathematical definition of optimal coarse graining for inhomogeneous systems with local structure.

	In concrete terms, our compression method is defined as follows. Consider a system of microscopic DOFs living on the graph $G$, defined, as usual, as the sets of vertices and edges. The physical space of configurations of the system living on the graph is described by a (large dimensional) random variable $\mathcal{X}$ distributed according to some joint probability distribution $P \left(\mathcal{X}\right)$. 
    The DOFs may exist either on the vertices, on the edges (as in the dimer model), or both.
 
    Let further $G = \bigcup_i V^i$ denote a decomposition of $G$ into a set of simply connected local subgraphs (patches). A coarse graining of a partition $\mathcal{X} = \bigcup_i \mathcal{V}^i$ of the physical configurations into new variables $\mathcal{X}'=\bigcup_i \mathcal{H}^i$ is then defined as a conditional probability distribution 
    \begin{equation}
        P(\mathcal{X}'|\mathcal{X})=\prod_i P_{\boldsymbol{\Lambda}^i}(\mathcal{H}^i|\mathcal{V}^i),
    \end{equation}
    where each factor is an individual coarse graining of block variable $\mathcal{V}^i \mapsto \mathcal{H}^i$. This will be a compression map by construction, so it monotonically reduces the entropy $H(\mathcal{H}^i) \leq H(\mathcal{V}^i)$. We describe a specific ansatz for such mappings below.

	We emphasise the distinction between the spatial patch $V^i$, and configurations of DOFs supported on this patch $\mathcal{V}^i$, which is a random variable. 
    The patch $V^i$ can be chosen as any local subgraph, for instance a topological ball defined using graph distance, or another set dictated by the structure of the problem, such as the tiles we used in the AB example (Fig.~2a).

    Algorithmically, the static graph structure allows us to define a constant indexing of the individual DOFs that is fixed across all real-space samples.
    Therefore, once we use the graph structure to define the subsystems $\mathcal{V}^i, \mathcal{E}^i$ we then can forget about the connectivity of the subgraphs $V^i, E^i$, and simply cast the DOFs into vectors:
    \begin{equation}
        \underbrace{\{\mathcal{V}^i_j\}}_{\text{set of DoFs}} \xrightarrow[\text{permutation}]{\text{fixed}} \underbrace{[\mathcal{V}^i_j]}_{\text{vector of DoFs}}.
    \end{equation}
    Therefore, we can use standard neural network (NN) architectures to handle the physical configurations on the graph.
    Note that would not be able to use this trick in the more general case of dynamic graphs, as it requires respecting the permutation invariance of the vertices.
    
    Maximisation of the real-space mutual information (RSMI) 
	\begin{equation*}\label{eq:midef}
        I(\mathcal{H}^i:\mathcal{E}^i)=\mathbb{E}_{P(\mathcal{H}^i, \mathcal{E}^i)}\left[ \log P(\mathcal{H}^i, \mathcal{E}^i) - \log P(\mathcal{H}^i)P(\mathcal{E}^i) \right]
	\end{equation*}
	between $\mathcal{H}^i$ and its distant environment $\mathcal{E}^i$ provides a variational principle for the coarse graining map $\boldsymbol{\Lambda}^i$.
    The objective depends on the coarse graining mapping via the compressed $\mathcal{H}^i$ variables: $I(\mathcal{H}^i:\mathcal{E}^i) \equiv I_{\boldsymbol{\Lambda}^i}(\mathcal{H}^i:\mathcal{E}^i)$. 
    The construction of the RSMI objective function enables distilling the most relevant large-scale features~\cite{PhysRevX.10.011037,gordon2020relevance}, as it tracks the correlations with a distant environment $\mathcal{E}^i$.
    Formal connections between this objective and the most relevant operators in critical lattice systems has recently been demonstrated both numerically and analytically~\cite{gokmen2021statistical, gordon2020relevance}.
    
	The compression is enforced by limiting the information capacity of $\mathcal{H}^i$ using a predetermined number of encoding bits (as we describe below), thereby directly implementing the rate constraint in the information bottleneck problem~\cite{gordon2020relevance, infbottle1}. Note that the compression is informed both of the underlying graph structure, and of the physics of the model living on it, which are encoded in the statistics of the configuration samples $(\mathcal{V}^i,\mathcal{E}^i)$.
	
    \subsection*{Estimation of mutual information}
	The computationally challenging RSMI variational principle can be efficiently implemented with differentiable lower bounds on mutual information~\cite{belghazi18a, Poole2019}. Such bounds are parametrised by a deep NN, which we call the \emph{neural critic} function (see below).
    The key idea behind this approach is that the estimation of mutual information $I(\mathcal{X}:\mathcal{Y})$ is converted to a classification task; where the \emph{neural critic}, $f(\mathcal{X},\mathcal{Y})$, is trained to distinguish so-called \emph{positive} and \emph{negative} sample pairs, which are sampled respectively from the true joint distribution $P(\mathcal{X},\mathcal{Y})$ and the product of marginals $P(\mathcal{X})P(\mathcal{Y})$~\cite{oord2019}.

    Some of the authors have recently used these techniques to develop a tractable implementation of the variational principle in Eq.~\ref{eq:rsmi} on regular lattices~\cite{gokmen2021statistical,gokmen2021phase}. 
    This is the RSMI-NE algorithm, where the neural critic $f$ is optimised simultaneously with a coarse graining ansatz $\boldsymbol{\Lambda}$ using stochastic gradient descent, as we describe below.

	In the present work we extended this compression framework for RG, and the RSMI-NE package to systems on arbitrary static graphs by casting the configurations into vectors according to the fixed coordinate system defined by the graph. The graph-enabled RSMI-NE code using the NetworkX backend~\cite{networkx_code} is available publicly~\cite{rsmine_code}.
	
    \subsection*{The coarse-graining and neural critic ansatze}
    We specify the coarse graining $P_{\boldsymbol{\Lambda}^i}(\mathcal{H}^i|\mathcal{V}^i)$ using an inner-product ansatz 
    \begin{equation}\label{eq:cg_ansatz}
        \mathcal{H}^i_k:=\tau \left(\boldsymbol{\Lambda}^i_{kj}  \mathcal{V}^i_j\right), 
    \end{equation}
    parametrised by a vector of linear NNs $\boldsymbol{\Lambda}^i=(\Lambda_k^i)$, and $\tau$, which is binary a discretisation map (\emph{e.g.} sign function).
    The index $k$ runs over the components of a coarse-grained variable, and the index $j$ refers to the spatial positions in region $\mathcal{V}$, where the indexing is defined with respect to a fixed labelling of vertices in the graph. Though we considered scalar DOFs in the AB dimer system, vector DOFs $\mathcal{V}^i_{jl}$ can also be handled by increasing the rank of the coarse graining filter by one: $\Lambda^i_{kjl}$. 
    In general, non-linear NN ansatze can also be used, and may be even necessary in certain cases, see for example Ref.~\cite{oppenheim2023machine}. 
    Our code supports such more general mappings that do not have this multilinear structure, and are instead parametrised by deep NNs.
    
    We implemented the binary mapping using the Gumbel-softmax trick~\cite{jang2017categoricalreparameterizationgumbelsoftmax}, which is an annealed relaxation of the Bernoulli distribution for $\mathcal{H}_k^i$. 
    This allows us to backpropagate through the discrete sampling process, and to train the coarse-graining filters using stochastic gradient descent, while ensuring a fixed rate of compression via discretisation.
    We used an annealing schedule exponential relaxation rate of $5\times 10^{-3}$, so that the variables are effectively binary at the end of the training.
    
    The critic function in the variational RSMI lower-bound is implemented using a separable architecture 
	\begin{equation}\label{eq:neural_critic}
		f(\mathcal{H},\mathcal{E}) = u(\mathcal{H})^{\rm T} v(\mathcal{E})
	\end{equation}
	where we used two-layer deep NNs for $u$ and $v$, with hidden dimension 16 and output dimension 8 (the hidden dimension is contracted in the product of the two networks). 
    
    \subsection*{Training details} 
    We trained the NNs using stochastic gradient descent with learning rate of $10^{-3}$ using 50000 sample dimer configurations, generated via the directed-loop Monte Carlo algorithm on the AB graph. 
    The total graph we considered contains 26177 nodes (the full graph is shown in the Supplementary Information). The sample dataset is supplied to the RSMI-NE algorithm in mini-batches of size 1000 and 120 epochs of the entire dataset.
	
    The coarse-grained block variable $\mathcal{V}$ at a given scale $\delta^s$ is defined on the $\sigma^s$ inflated tiles $V$ shown with different colours in Fig.~2a. The corresponding environment regions $E$, are defined as a shell with radius given by a fixed graph-distance from the centre of $V$. In particular for $\delta^2$, $E$ is defined by an inner radius $L_{E_{\rm in}}=9$ and outer radius $L_{E_{\rm out}}=24$, whereas for $\delta^4$ we used $L_{E_{\rm in}}=40$, $L_{E_{\rm out}}=64$, as shown in Supplementary Fig.~2. Examples of the corresponding  $\sigma^{-4}$ coarse-graining filters are shown in Supplementary Fig.~3.

    \section*{Data availability}
    The data generated during the course of this study have been deposited in the \emph{Figshare} repository at \url{https://doi.org/10.6084/m9.figshare.27245481} (Ref.~\cite{rsmi_ab_data_2024}).

    \section*{Code availability}
    The RSMI-NE software used in this study is available as an open-source repository in the \emph{Zenodo} repository linked in Ref.~\cite{rsmine_code} and \url{https://github.com/RSMI-NE/RSMI-NE}.
    
    \bibliography{refs}
	
	\section*{Acknowledgements}
    D.E.G. gratefully acknowledges support from the Simons Foundation and NSF through the National Institute for Theory and Mathematics in Biology.
	D.E.G., and S.D.H. acknowledge financial support from the Swiss National Science Foundation Grant No. 182240. S.B. acknowledges support by the European Research Council under the European Union Horizon 2020 Research and Innovation Programme via Grant Agreement No. 804213-TMCS. Z.R. acknowledges support from ISF grant 2250/19. F.F. acknowledges support from EPSRC Grant No. EP/X012239/1. M.K.-J. gratefully acknowledges financial support from the European Union’s Horizon 2020 programme under Marie Sklodowska-Curie Grant Agreement No. 896004 (COMPLEX ML).

    \section*{Author Contributions}
    D.E.G., S.B., S.D.H., Z.R., F.F. and M.K.-J. designed the research, performed the research and wrote the manuscript.

    \section*{Competing Interests}
    The authors declare no competing interests.

\ifthenelse{\boolean{captionSection}}{
    \section*{Figure Legends}

    \subsection{Figure 1}
    {{\bf Schematic for constructing collective degrees of freedom in inhomogeneous systems.} Distinct systems like tissues (left, in green) and colloidal suspensions (right, in blue) can be abstracted into a set of vector degrees of freedom $\mathcal{V}^i$ (indicated by stacks of squares, $i=1,2,3$) living on an irregular graph with local structure. The final component of each vector is shown by a coloured box to indicate potentially different types of internal degree of freedom, unique to each sub-system. To derive a compressed representation of such systems, it is essential to tailor the coarse graining transformation $\boldsymbol{\Lambda}^i$ for each local neighbourhood $i$. This is achieved by an information theoretic variational principle, where $\boldsymbol{\Lambda}^i: \mathcal{V}^i \mapsto \mathcal{H}^i$ maximises the mutual information $I\left(\mathcal{H}^i:\mathcal{E}^i\right)$. This allows the compressed variables $\mathcal{H}^i$ to capture the emergent long-range physics according to the statistics of the surrounding distant environment $\mathcal{E}^i$. Local optimisation can produce compressed variables with varying cardinality across the system, here illustrated by vectors $\mathcal{H}^i$ with varying numbers of components. The connectivity of the emergent \emph{supergraph} is determined through the correlations of the new variables.}

    \subsection{Figure 2}
    {{\bf Self-similarity of the Ammann-Beenker tiling, and the coarse graining blocks.} {\bf a} A microscopic dimer configuration on the AB tiling's edges, with an overlaid AB \emph{super-quasilattice}, self-similar to the microscopic one. The effective degree of freedom at a supervertex with valence $n$ will be obtained by coarse graining the dimer configuration in the surrounding polygon tile $V^n$. In total there are 4 classes of such polygons, here shown in green, blue, red and yellow for $n=8,3,4,5$, respectively. The shape of the block tile is dictated by the valence $n$ of the central supervertex in matching colour. {\bf b} The inflation (deflation) $\sigma^{2(-2)}$ of the elementary rhombi and squares generating the tiling, with parts of the polygonal domains indicated in colour. Coarse graining all such polygonal patches executes a deflation $\sigma^{-2}$ of the original AB quasilattice, yielding the super-quasilattice shown.}

    \subsection{Figure 3}
    {{\bf Finding collective clock variables.} {\bf a} Coarse graining transformation $\boldsymbol{\Lambda}$ compressing Monte Carlo configurations $\mathcal{V}$ into bitstrings $\mathcal{H}$ on supervertices of the $\sigma^{-2}$ deflated AB tiling. Each bit $\mathcal{H}_k$ is decided by the sign of the linear transformation $\Lambda_k \cdot \mathcal{V}$. {\bf b}({\bf f}) The length of the bitstring $\mathcal{H}^{8(3)}$ is determined by the saturation point (shown in green) of mutual information at 4 (2) bits at 8- (3-)supervertices. {\bf c}, {\bf g} The respective optimal filters $\boldsymbol{\Lambda}^8$ and $\boldsymbol{\Lambda}^3$ carry a representation of the local spatial symmetries of corresponding supervertices, namely $\mathsf{C}_8$ and mirror. {\bf d}({\bf h}) The probability distributions $P(\mathcal{H}^{8(3)})$ occupy the space of codes sparsely, and form abstract $\mathbb{Z}_{8(3)}$ clock variables. {\bf e} In particular, $\mathcal{H}^8$ forms a closed 8-loop, where each state has exactly two neighbours with Hamming-distance 1. {\bf i} The representations of the local symmetries on filters induce transitions between adjacent clock-states, enabling the identification of abstract clock-states with spatial directions along the links of the quasiperiodic lattice.
    }

    \subsection{Figure 4}
    {{\bf Emergent dimer exclusion rule and self-similar dimer-dimer correlations across scales.} {\bf a} The probability distribution of microscopic (\emph{i.e.~}$\delta^0$) dimers (in greyscale) on an AB patch, conditioned on one of the links (in orange) hosting a dimer. {\bf b}, {\bf c} First two columns: the probabilities $P(\mathcal{H}^8|\mathcal{H}^3)$ of the emergent clock variables on the $\delta^{2}$ and $\delta^{4}$ super-quasilattice (in greyscale), conditioned on two distinct states of one of the 3-clocks (in orange). The third column shows distributions $P(\mathcal{H}^3|\mathcal{H}^8)$ conditioned on a state of the central 8-clock. Binding of adjacent clock variables into super-dimers obeying dimer exclusion constraints is revealed by sharply peaked conditional distributions. The effective super-dimers reproduce also longer-range dimer-dimer correlations at both $\delta^2$ and $\delta^4$ scales. {\bf d}, {\bf e} Examples of optimal coarse-graining filters producing the central 8-state clock variable at scales $\delta^{2}$ and $\delta^{4}$. The latter comprises $2760$ microscopic links.
    }
}{}

\ifthenelse{\boolean{preprint}}{
    \newpage
    \section*{Supplementary information: Compression theory for inhomogeneous systems}
    
    \section{The odd scale transformations of dimer coverings on the AB tiling}

	Our analysis of the coarse graining transformations of the dimer model on the AB tiling \emph{did not} provide evidence for a discrete scale invariant description in terms of super-dimer variables under all rescalings, but only for \emph{even} order ones (\emph{i.e.}~under deflations $\sigma^{-2k}$, for $k\in\mathbb{N}_0$). This is in contrast to the AB tiling itself (\emph{i.e.~}just the AB quasilattice), which is invariant under any order of deflation.

    \begin{figure}[ht!]
		\centering
		\includegraphics[width=0.5\textwidth]{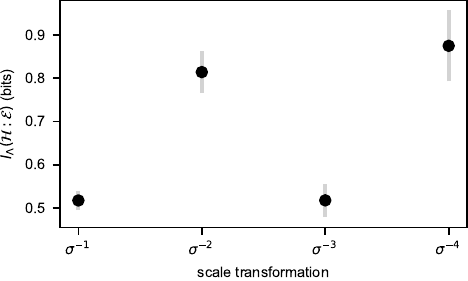}
		\caption{Mutual information across different scale transformations.
			The maximal MI for the coarse graining at a 3-supervertex. For odd order rescaling transformations $\sigma^{-1,-3}$, the information attained by the compression is systematically lower compared to the even ones $\sigma^{-2,-4}$. The error bars show the standard deviation of the mutual information value over 5 independent runs of the RSMI-NE optimisation procedure.}\label{fig:figS1}
	\end{figure}
	
	In addition, our method finds quantitatively and qualitatively distinct behaviour at odd orders $\sigma^{-1}$ and $\sigma^{-3}$. The maximal mutual information $I_\Lambda(\mathcal{H}:\mathcal{E})$ attained for the coarse graining at a 3-supervertex is non-monotonic, exhibiting, within error, two distinct values characterizing the even and odd scales (with the odd scales' information reduced to almost half), as shown in Supplementary Fig.~\ref{fig:figS1}. 

	Furthermore, the optimised coarse graining does not yield a well-defined three-state clock variable at odd scales. Indeed, for even scales the optimisation robustly yields a well-defined set of three clusters (corresponding to the three clock states) even in the distribution of pre-activations $\mathbf{\Lambda} \cdot \mathcal{V}$, while at odd scales the distribution of pre-activations lacks any such clear structure. We emphasize that this is \emph{not} an optimization error: computationally, $\sigma^{-4}$ coarse graining is a more challenging problem than $\sigma^{-3}$, due to a significantly larger number of degrees of freedom involved.

\clearpage
\section{Coarse-graining blocks and filters for $\sigma^{-4}$ deflation} 

	\begin{figure*}[h!]
		\centering
		\includegraphics[width=\textwidth]{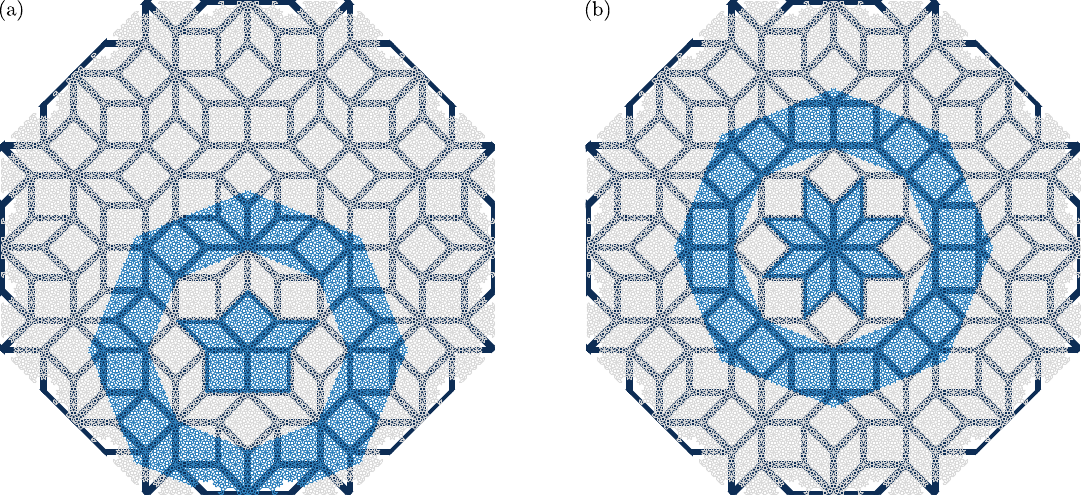}
		\caption{Block and environment regions. Highlighted in blue are examples of the coarse-graining blocks $V$, and their annular environment regions $E$ used for the 
			3- (a) and 8-vertices (b) at the largest scale considered (\emph{i.e.~}$\delta^4$). The microscopic quasilattice, and the $\sigma^{-2}$ super-quasilattice are shown.
			The centers of the `kite' and `star' shaped regions $V$ are at the 8-vertices whose positions form the $\sigma^{-4}$ super-quasilattice.}\label{fig:figS2}
	\end{figure*}

	\begin{figure*}[h!]
		\centering
		\includegraphics[width=\textwidth]{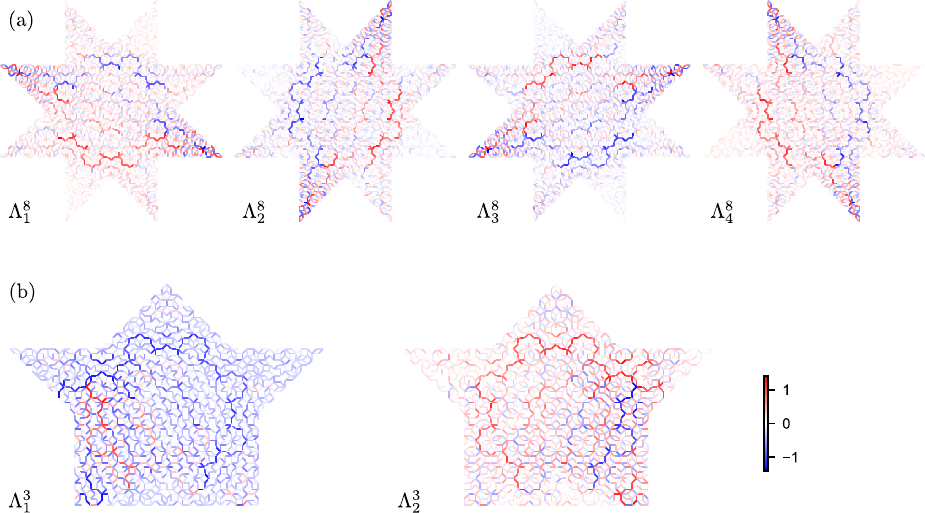}
		\caption{Optimal $\sigma^{-4}$ coarse-graining transformations.
			(a) 8-supervertex filters. (b) 3-supervertex filters.}\label{fig:figS3}
	\end{figure*}

\section{RSMI compression on the AB-tiling with distance based blocks}

	In this section we demonstrate that our results are robust to different choices of the coarse-graining blocks. 
	We repeat the same analysis we did for the dimer model on the AB tiling, but instead of the elementary AB tiles, we define the blocks using the graph distance. 
	The optimal coarse-graining transformations at the 8-vertices are shown in Supplementary Fig.~\ref{fig:dist_filters}, where length scale $L_V$ is chosen to correspond to a $\sigma^{-4}$ deflation of the AB tiling. 
	As in the main text, we find that the emergent DOFs are discrete clock variables, with number of states equal to the connectivity of the supervertex.

	\begin{figure*}[ht]
		\centering
		\includegraphics[width=\textwidth]{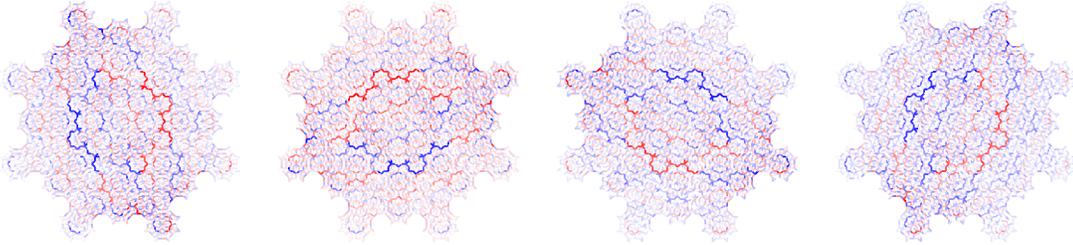}
		\caption{Optimal $\sigma^{-4}$ coarse-graining transformations at the 8-vertices of the AB tiling, where the coarse-grained blocks are defined using the graph distance instead of the elementary AB tiles in Supplementary Fig.~\ref{fig:figS3}.}\label{fig:dist_filters}
	\end{figure*}

	We probe the correlations of the clock variables using conditional probabilities and find that the emergent dimer exclusion is preserved, as shown in Supplementary Fig.~\ref{fig:dist_correlations}.
	Conditioned on an adjacent 3-clock $\mathcal{H}^3$ pointing to the centre, the probability $P(\mathcal{H}^8|\mathcal{H}^3)$ of the central 8-clock peaks at the state corresponding to the same direction. This indicates that the emergent DOFs are still paired with one and only one of their neighbours into emergent super-dimers on the edges of the super-quasilattice, as in the case of the elementary AB tiles.

	\begin{figure}[ht]
		\centering
		\includegraphics[width=8cm]{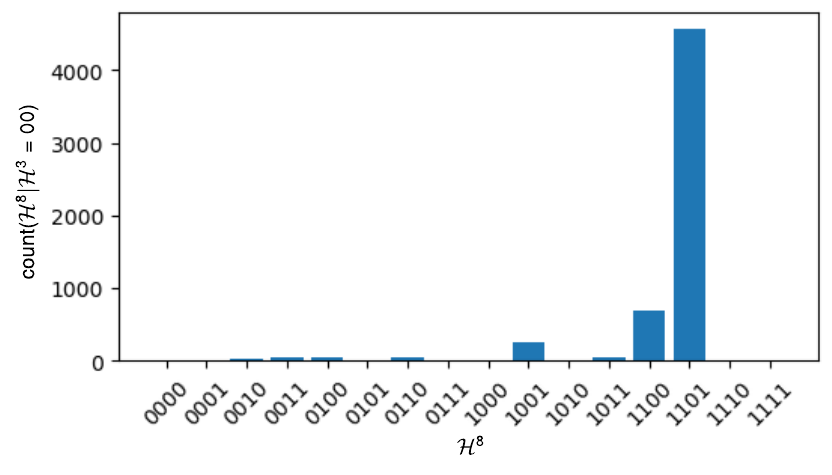}
		\caption{The emergent dimer exclusion for resulting clock variables obtained using the coarse-grained blocks defined using graph distance. Conditioned on an adjacent 3-clock $\mathcal{H}^3$ pointing to the centre (\emph{i.e.} $\mathcal{H}^3={\tt 00}$), the probability $P(\mathcal{H}^8|\mathcal{H}^3)$ of the central 8-clock peaks at the state corresponding to the same direction.} \label{fig:dist_correlations}
	\end{figure}

	\section{Antiferromagnetic Ising model on nearly bipartite random graphs}  
	\begin{figure*}[h]
		\includegraphics[width=0.8\textwidth]{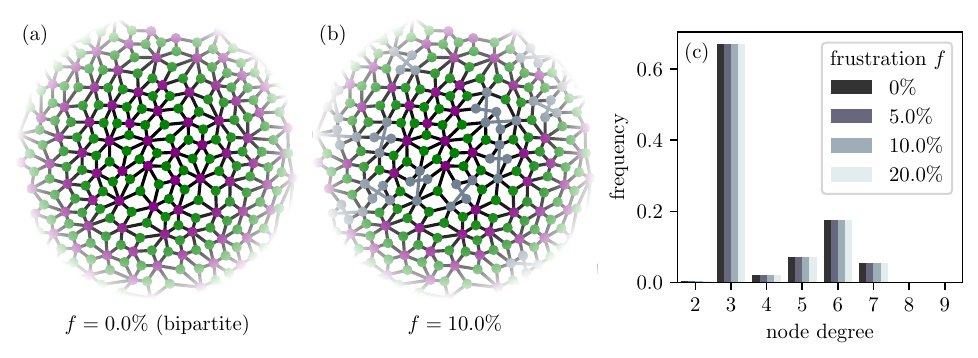}
		\caption{
			Random graph ensemble with tunable frustration. (a) First, we construct random planar bipartite graphs by decorating the triangulation of a densely packed set of disks (purple nodes) with additional nodes at the triangle centres (green nodes), and placing the edges to connect the new node to the triangle corners.  (b) Next, we introduce frustration by breaking the bipartite structure in a controlled way. This is done by deleting and crossing the opposite edges in a fraction $f$ of the rhombic faces. (c) Through this process, we maintain fixed node-coordination statistics while varying the level of frustration.}
		\label{fig:random_graph}
	\end{figure*}
	In this section, we present an application of the RSMI-NE algorithm to frustrated antiferromagnets on random graphs, which is closely related to graph bisection and colouring problems in combinatorial optimisation and theoretical computer science~\cite{random_bipartitioning}.
	Our results demonstrate the versatility of this method across a broader class of quenched disordered systems and its applicability to more generic aperiodic graph topologies where there is no canonical choice for the block shapes.

	\paragraph{Model --} We consider the antiferromagnetic (AFM) Ising model on a random graph $G$. The Hamiltonian is given by
	\begin{equation}\label{eq:rand_afm}
		E[\bm{s}] = -J \sum_{(i,j)\in e_G} s_i s_j,
	\end{equation}
	where $J<0$, $s_i = \pm 1$ are Ising spins, and the sum runs over all edges $e_G$ of the graph $G$. 

	The problem of finding the ground state of this system is equivalent to dividing the set of graph nodes into two subsets, such that the number of edges across the two subsets is maximised, and those within the subsets is minimised. 
    The ground state energy $E_0$ then gives the \emph{maximum cut} $|E_0|$.
    This prototypical combinatorial optimisation problem is known to be NP-complete on generic graphs~\cite{GAREY1976237, garey1979computers, Karp1972}. However, for bipartite graphs, it reduces to the trivial two-colouring problem, where there is a maximum cut that is unique up to permutation of the partitions (or a global $\mathbb{Z}_2$ spin flip). Moving away from bipartiteness, the ground-state manifold likely becomes highly degenerate due to frustration~\cite{random_bipartitioning}.

    As a more tractable special case of this difficult problem, we narrow our focus to graphs that deviate from the bipartite structure by a small controlled amount. 
This case relates closely to the planted-partition model~\cite{BOLLOBÁS_SCOTT_2004}, where nodes are divided into predetermined partitions, with edges placed across partitions with a small probability and within partitions with a higher probability. 
Planted models are useful for benchmarking improved partitioning algorithms, since the max-cut is significantly larger than the expected cut of a randomised partitioning, \emph{cf.} generic dense random graphs~\cite{goemans_williamson}. 
Furthermore, since the planted partition is expected to be close to the maximal cut, it is easier to verify the solution.
    
	We will specifically consider a particular ensemble of random graphs generated via the following two-step procedure:
	\begin{enumerate}
		\item We first construct a planar graph by decorating the triangulation of a 2D random point set with additional nodes at the triangle centres, and placing the edges $e_G$ to connect the new node to the triangle corners. This results in a graph with a (planted) bipartite structure, as shown in Supplementary Fig.~\ref{fig:random_graph}a.
		\item Following this, we introduce frustration by deliberately disrupting the bipartite structure. This involves selectively removing and crossing pairs of opposite edges within a portion $f$ of the rhombic faces, as illustrated in Supplementary Fig.~\ref{fig:random_graph}b.
		Through this process, we can adjust the level of frustration while maintaining fixed node-coordination statistics, as evidenced in Supplementary Fig.~\ref{fig:random_graph}c.
		This way we can isolate the effects of frustration on the collective DOFs and their correlations without changing the effective dimensionality of the system. 
	\end{enumerate}

    On perfectly planar graphs, the duality of the max-cut problem to the route-inspection problem leads to an exact polynomial-time algorithm~\cite{route_inspection_planar_maxcut}. 
    Our random graph ensemble is non-planar.
    The above bond crossing procedure (2.) allows us to tune the hardness of the problem.
    In contrast to an arbitrary procedure for departing from exact-planarity, this maintains the size of the maximum cut to be significantly larger than half the number of edges.
    These features make this ensemble suitable to demonstrate and evaluate the performance of RSMI-NE compression in extracting relevant observables in aperiodic and amorphous geometries.

    The current state-of-the-art approximation algorithm is a polynomial-time semidefinite programming method due to Goemans and Williamson.
    There exist other heuristic approaches such as greedy local one-exchange algorithm, which can perform well in specific ensembles.
    However, the convergence time of the latter often has poor scaling and it does not provide approximation guarantees as it can get stuck in local minima. 

	\paragraph{Results --} 

	\begin{figure*}[t]
		\centering
		\includegraphics[width=0.85\textwidth]{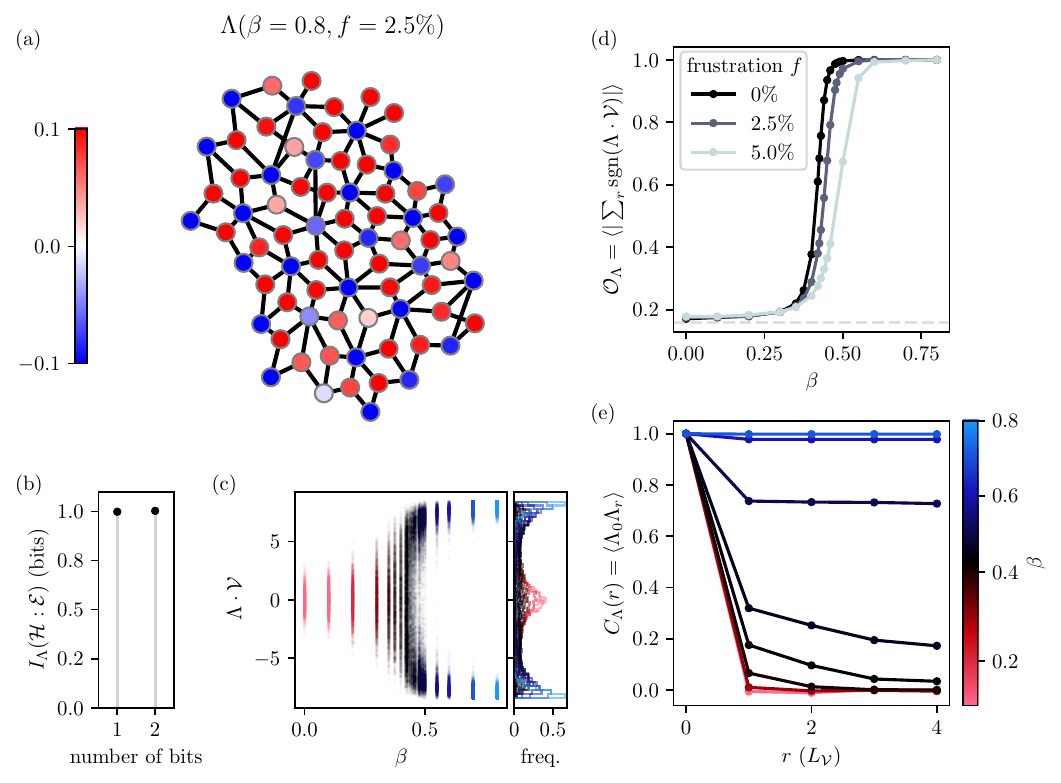}
		\caption{(a) The optimal coarse-graining filter $\Lambda$ at frustration $f=2.5\%$ and inverse temperature $\beta=0.8$. (b) We use a 1-bit coarse-grained variable $\mathcal{H}={\rm sgn}(\Lambda \cdot \mathcal{V})$ as the mutual information value saturates using a single filter component. (c) The distribution of the pre-activation $\Lambda \cdot \mathcal{V}$ transitions from unimodal to bimodal as a function of $\beta$, indicating spontaneous symmetry breaking. (d) The individually optimised local coarse-graining transformations can be combined into a global order parameter $\mathcal{O}_\Lambda$ for finite $f$. (e) The correlation function of the coarse-grained variable $\mathcal{H}$ exhibits long-range order at low-temperature for finite frustration (here $f=2.5\%$).}\label{fig:CG_analysis}
	\end{figure*}

    We treat this problem as an equilibrium statistical physics model of a magnet, exhibiting critical phenomena.
    For the bipartite case ($f=0\%$), it is expected that this system undergoes a second-order phase transition into a long-range ordered AFM phase at a sufficiently low temperature. It is then natural to ask whether this long-range order persists under finite frustration.
    Answering this question requires constructing an AFM order parameter on random graphs: a task which becomes challenging when the bipartitioning is neither known nor unique.

    To address these questions, we apply the RSMI-NE compression on Monte Carlo configurations generated using the Wolff cluster algorithm~\cite{PhysRevLett.62.361}.
	We construct the block $\mathcal{V}=[s_i]_{i \in V}$ and the environment $\mathcal{E}=[s_i]_{i \in E}$ using sets of nodes $V,E$ partitioned via the graph distance. In particular we set $L_V=5$, $L_B=8$, and $L_{E_{\rm out}}-L_{E_{\rm in}}=8$.
	Like in Fig.~\ref{fig:fig3} in the main text, the number of filters is determined by the plateauing of mutual information, which, in this case, is 1 bit (Supplementary Fig.~\ref{fig:CG_analysis}b).

    The optimal coarse-graining filter successfully identifies the correct planted bipartitioning (Supplementary Fig.~\ref{fig:CG_analysis}a) on the block $\mathcal{V}$ at frustration $f\leq 7.5\%$. (We benchmark the bipartitioning of the RSMI-compression against other standard algorithms below.)
	In turn, the emergent DOF $\mathcal{H}={\rm sgn}(\Lambda \cdot \mathcal{V})$ is a local order parameter for the AFM order. 
	This can be seen in Supplementary Fig.~\ref{fig:CG_analysis}c, where the distribution of the pre-activation $\Lambda \cdot \mathcal{V}$ transitions from unimodal to bimodal as a function $\beta$.

	To check if the long-range order persists under finite frustration, we need to construct a global order parameter. 
	We can do this by combining the individually optimised local coarse-graining transformations at blocks $\mathcal{V}^i$ that span the entire graph: 
	\begin{equation}\label{eq:afm_rsmi_op}
		\mathcal{O}_\Lambda = \left \langle \left\|\sum_i {\rm sgn}(\Lambda^i \cdot \mathcal{V}^i) \right\| \right \rangle = \left \langle \left\|\sum_i \mathcal{H}^i \right\| \right \rangle,
	\end{equation}
    Since the mutual information is invariant under a sign flip $\mathcal{H}\mapsto -\mathcal{H}$, constructing $\mathcal{O}$ requires fixing the gauge for independently trained $\Lambda^i$. 
    We do this by a `stitching' procedure, where we take slightly overlapped regions $\mathcal{V}^i$ and match the sign of the filters $\Lambda^i$ in the overlapping region by flipping the sign of the entire filter, when necessary.
 
    Computing $\mathcal{O}_\Lambda(\beta)$ (Supplementary Fig.~\ref{fig:CG_analysis}d), we find that the system perfectly `magnetises' at low temperatures for $f\leq 7.5\%$, while the sole effect of frustration is to shift the critical temperature to lower values.
	Finally, we can probe the correlation function of the coarse-grained variable $\mathcal{H}$, which exhibits long-range order at $\beta > 0.5$ for finite frustration, as shown in Supplementary Fig.~\ref{fig:CG_analysis}e.

    Combined with the above mutual information gauge stitching procedure, the RSMI-NE algorithm thus assisted us in constructing a global order parameter for the antiferromagnetic phase on random graphs.
    This enabled us to study the effects of frustration on long-range order without prior knowledge of the bipartitioning, and conclude that finite but small frustration $f\leq 7.5\%$ is irrelevant for large-scale physics.

       \begin{figure}[h!]
		\centering
		\includegraphics[width=0.9\textwidth]{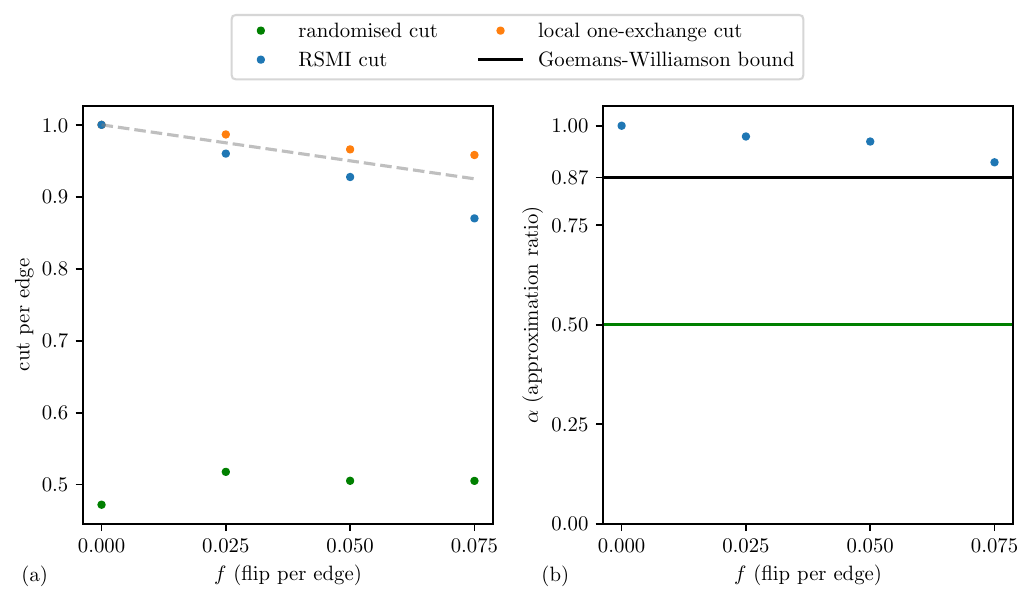}
		\caption{Benchmarking RSMI-NE in the max-cut problem. 
            (a) The size of the cut (per edge) for different algorithms. The RSMI-NE cut (blue) gets close to the optimal solution found by the greedy local one-exchange strategy (orange). The max-cut is significantly larger than the randomised cut (green) of $0.5$ per edge.
            The gray dashed line is $1-f$.
            (b) Approximation ratio for RSMI-NE is higher than the performance guarantee of $\alpha\geq 0.87$ for the Goemans-Williamson algorithm (solid black line).
        }\label{fig:fig_maxcutbench}
	\end{figure}

    \paragraph{Benchmarking.} Since max-cut is NP-hard, exact algorithms, such as branch and bound, are infeasible for the graphs we consider with several hundreds of nodes.
    Moreover, this problem is also APX-hard~\cite{maxcut_apx_hard}, \emph{i.e.} it does not have a polynomial-time approximation scheme that is guaranteed to converge to the global extrema.
    Nevertheless, there exist approximation algorithms that are guaranteed find a cut that is at least a certain proportion of the optimal value.

    A commonly used metric to measure the performance of an approximation scheme is the \emph{approximation ratio}
    \begin{equation}
        \alpha(E) = \frac{E}{E_0},
    \end{equation}
    so that $\alpha=1$ when the cut is indeed maximal.

    We first provide a comparison of RSMI-NE with randomised partitioning and greedy one-exchange local search algorithms.
    For this benchmarking, we consider instances from the above random graph ensemble in a range of values for $0\%<f<7.5\%$, each graph consisting of 1084 nodes in total.
    As shown in Supplementary Fig.~\ref{fig:fig_maxcutbench}a, the RSMI cut fraction is comparable to the one-exchange algorithm, which is expected to perform well in the regime close to exact bipartiteness and planarity (in the shown instances, it indeed finds the global optimum).
    Note however, that the convergence time of iterated local search exceeded the runtime of RSMI-NE significantly and for the graphs we considered (see Supplementary Fig.~\ref{fig:maxcut_runtime}).

    There are a few factors to consider regarding the scaling of RSMI-NE.
    Firstly, the runtime is mostly dominated by the deep neural network critic function $f$ (Eq.~\ref{eq:neural_critic}), whose task is to estimate the mutual information.
    Since the environment $\mathcal{E}$ comprises complementary set of nodes to $\mathcal{V}$, the forward passes of $f$ involve the combined $(\mathcal{V}, \mathcal{E})$ graphs that are always roughly the same size.
    Therefore, this part of the computation should not scale with the number of nodes in $\mathcal{V}$.
    The main contributor of the scaling is the increasing number of training iterations till reaching 1-bit, followed by a weaker contribution from the forward passes of the compression map $\Lambda\cdot \mathcal{V}$, which is a linear neural network (computationally much lighter than $f$).
    Note also that the runtime of the Monte Carlo algorithm is included to RSMI-NE in Supplementary Fig.~\ref{fig:maxcut_runtime}. 
    While the Wolff algorithm has linear scaling with the number of nodes, we obtained the $(\mathcal{V}, \mathcal{E})$ configurations from the same one large graph of 1804 nodes. Therefore, the MC runtime amounts to a constant offset of 2.36 seconds for generating 10k samples on this graph.

    One major drawback of the one-exchange algorithm is the lack of a formal guarantee to find the globally optimal solution as it can get stuck in local minima.
    A reference benchmark \emph{with} an approximation guarantee is the Goemans-Williamson (GW) algorithm~\cite{goemans_williamson}. 
    It has the best known approximation guarantee, $\alpha\geq0.87$.
    We observed that the approximation ratio of RSMI exceeds the worst-case scenario of the GW algorithm (Supplementary Fig.~\ref{fig:fig_maxcutbench}b).
    Notice though, that the performance of RSMI-NE degrades with increasing $f$, and eventually drops below the GW bound.

    \begin{figure}[h!]
        \centering
        \includegraphics[width=3.9in]{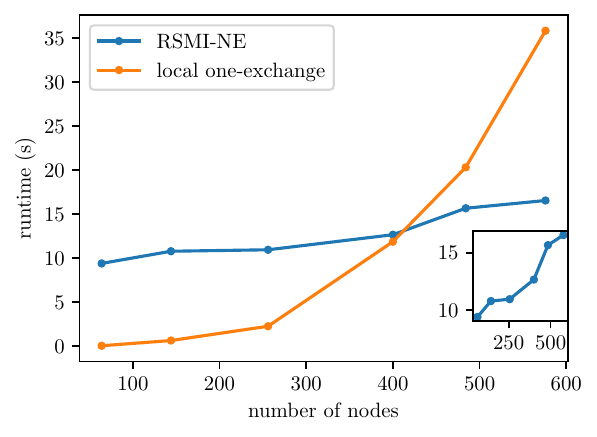}
        \caption{
        Runtime scaling of the local one-exchange algorithm and the RSMI-NE algorithm with the number of nodes in the graph.
        The local search algorithm runs a greedy one-exchange strategy until no improvement can be made.
        The RSMI-NE algorithm is ran until it achieves 1-bit of mutual information, which signals when the compression map captures the long-range order. The sudden increase in runtime for RSMI-NE around 500 nodes (see the inset) is due to the increased number of epochs to reach 1-bit.
        }
        \label{fig:maxcut_runtime}
    \end{figure}

    We emphasise that RSMI-NE had not been hard-coded to solve the max-cut problem, but it has \emph{identified} this as the most relevant question, and extracted the AFM order parameter.
    While tuning up $f$, one needs to consider the possibility that this system undergoes a qualitative change, whereby antiferromagnetism is destroyed.
    In this case, the RSMI compression map may start to track a \emph{new} slow observable, whose mixing with the AFM order in the intermediate regime might be the reason for the degrading the max-cut performance.
    These questions concerning the extended phase diagram of Eq.~\ref{eq:rand_afm} go beyond the scope of this appendix and we leave this interesting possibility for a future study.

    Since the RSMI-NE compression selects the bipartitioning as the correct relevant observable, the present results suggest that one can indeed directly use one of the existing bipartitioning algorithms (\emph{e.g.} the Goemans-Williamson algorithm, see below) as a \emph{computational order paremeter}~\cite{weinstein2024computationalphasetransitionstwodimensional} to detect the long-range ordering in this system, also without employing any coarse-graining procedure.
    In Supplementary Fig.~\ref{fig:algo_magnetisation}a, we plot the temperature dependence of such an AFM order parameter constructed according to the one-exchange bipartitioning ($f=5\%$):
    \begin{equation}
        m = \sum_{i \in {\rm set 1}} s_i - \sum_{i \in {\rm set 2}} s_i.
    \end{equation}
    We also provide a coarse grained version of this order parameter in Supplementary Fig.~\ref{fig:algo_magnetisation}b, defined analogously to Eq.~\ref{eq:afm_rsmi_op}. Again we construct Kadanoff block spins on blocks of radius $L_V = 5$, which leads to a $5\times 5$ coarse grained lattice.
    Note that the dashed line indicates the exact value for the non-vanishing absolute magnetisation density of a perfect paramagnet on a $5\times 5$ lattice. 
    Both versions indicate a phase transition around $\beta\approx 0.45$.
    It is unsurprising that the behaviour of this computational observable is almost identical to Fig.~\ref{fig:CG_analysis}d, as the RSMI-NE achieved an approximation ratio $\alpha\approx 0.95$ for the max-cut (Supplementary Fig.~\ref{fig:fig_maxcutbench}b).

    \begin{figure}
        \centering
        \includegraphics[width=\textwidth]{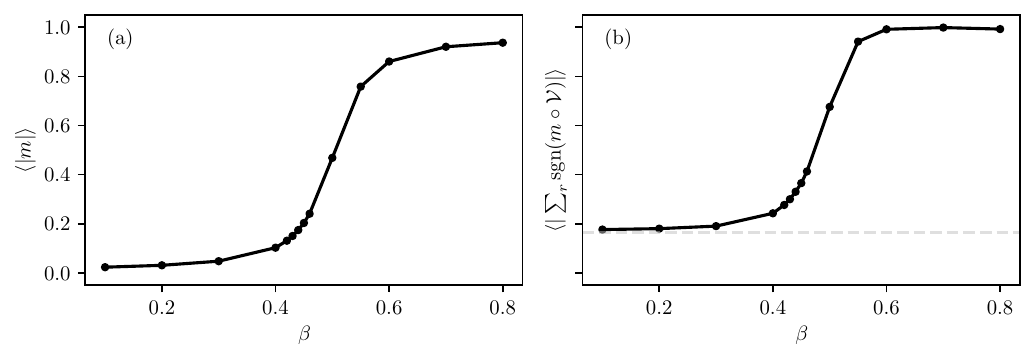}
        \caption{(a) Computational AFM order parameter constructed via the one-exchange local greedy search algorithm. (b) Coarse grained version.
        The temperature dependence is shown for the same $f=5\%$ random graph as for Supplementary Fig.~\ref{fig:CG_analysis}d.}
        \label{fig:algo_magnetisation}
    \end{figure}

\bibliography{refs}
}{}

\end{document}